%% file: paper.tex
\PassOptionsToPackage{dvipsnames,svgnames,x11names}{xcolor} 

\documentclass[sigplan,screen,authorversion]{acmart}

\usepackage{amsmath}
\usepackage{amsthm}
\usepackage{subcaption}
\usepackage{supertabular}
\usepackage{stmaryrd}
\usepackage{xcolor}
\usepackage{multirow}
\usepackage{calc}
\usepackage[T1]{fontenc}
\usepackage{graphicx}
\usepackage{xspace}
\usepackage{dashrule}  
\usepackage{stackrel}
\usepackage{enumerate}
\usepackage{framed}
\usepackage{bussproofs}
\usepackage{mdframed}  
\usepackage{comment}
\usepackage{ifthen}
\usepackage{pdftexcmds}
\usepackage{mathpartir}
\usepackage{fancyvrb}
\usepackage{booktabs}
\usepackage{tikz}
\usepackage{tikz-cd}
\usepackage[zi4]{inconsolata-slanted} 
\usepackage{listings}
\usepackage{url}
\usepackage{float}
\usepackage{subfiles}

\usetikzlibrary{calc}
\tikzset{curve/.style={settings={#1},to path={(\tikztostart)
    .. controls ($(\tikztostart)!\pv{pos}!(\tikztotarget)!\pv{height}!270:(\tikztotarget)$)
    and ($(\tikztostart)!1-\pv{pos}!(\tikztotarget)!\pv{height}!270:(\tikztotarget)$)
    .. (\tikztotarget)\tikztonodes}},
    settings/.code={\tikzset{quiver/.cd,#1}
        \def\pv##1{\pgfkeysvalueof{/tikz/quiver/##1}}},
    quiver/.cd,pos/.initial=0.35,height/.initial=0}

\usepackage{lstparams}
\usepackage{lstcoq}

\usepackage[capitalize, nameinlink]{cleveref}

\usepackage{siunitx}
\newcommand\cdh[1]{\lstinline[language=Haskell,breakatwhitespace]{#1}}

\newcommand\cdc[1]{\lstinline[language=Coq,breakatwhitespace]{#1}}
\newcommand\zcdc[1]{\let\par\endgraf\cdc{#1}}
\newcommand\zcdh[1]{\let\par\endgraf\cdh{#1}}

\newcommand\etc{\textit{etc.}}
\newcommand\ie{\textit{i.e.,\ }}
\newcommand\eg{\textit{e.g.,\ }}
\newcommand\vs{\textit{vs.\ }}

\newboolean{extended}

\usepackage{draft} 
\newnote{scw}{blue}       
\newnote{ly}{blue}      
\newnote{lx}{olive}      
\newnote{lai}{cyan}    
\newnote{gsv}{cyan} 
\newnote{gs}{orange}

\input{directives.tex}

\setcopyright{rightsretained}
\acmPrice{}
\acmYear{2025}
\copyrightyear{2025}
\acmConference[Haskell 2025]{the ACM SIGPLAN Haskell Symposium 2025}{Oct. 16--17}{Singapore}

\begin{document}
\ifextended
\title{Freer Arrows and Why You Need Them in Haskell (Extended Version)}
\else
\title{Freer Arrows and Why You Need Them in Haskell} 
\fi

\author{Grant VanDomelen}
\email{gsv@pdx.edu}
\orcid{0009-0002-2923-6415}
\affiliation{
  \institution{Portland State University}
  \country{USA}
}

\author{Gan Shen}
\email{gshen42@ucsc.edu}
\affiliation{
  \institution{University of California, Santa Cruz}
  \country{USA}
}

\author{Lindsey Kuper}
\email{lkuper@ucsc.edu}
\affiliation{
  \institution{University of California, Santa Cruz}
  \country{USA}
}

\author{Yao Li}
\email{liyao@pdx.edu}
\orcid{0000-0001-8720-883X}
\affiliation{
  \institution{Portland State University}
  \country{USA}
}

\begin{abstract}
\subfile{abstract.txt}
\end{abstract}

\maketitle

\bibliographystyle{ACM-Reference-Format}

\subfile{introduction}

\subfile{background}

\subfile{freer-arrow}

\subfile{freer-choice}

\subfile{haschor}

\subfile{discussion}

\subfile{related-work}

\subfile{conclusion}

\bibliography{ref, yao}

\subfile{appendix}

\end{document}

%% file: directives.tex
\draftfalse
\extendedfalse

%% file: abstract.txt
Freer monads are a useful structure commonly used in various domains due to their expressiveness. However, a known issue with freer monads is that they are not amenable to static analysis. This paper explores freer arrows, a relatively expressive structure that is amenable to static analysis. We propose several variants of freer arrows. We conduct a case study on choreographic programming to demonstrate the usefulness of freer arrows in Haskell.

%% file: introduction.tex
\section{Introduction}\label{sec:intro}

We love monads~\citep{moggi-monad, wadler-monad}. We use them all the time. Why
not? They are general. They are abstract. They are expressive. They allow us to
do diverse things under the same interface.

For this reason, it should be no surprise that structures like freer
monads~\citep{freer} and their variants~\cite{itree, mcbride-free, one-monad,
  steelcore, resumption, delay} have been used in various domains including
choreographic programming~\citep{haschor}, concurrency~\citep{haxl}, algebraic
effects~\citep{polysemy, freer-simple, fused-effects},
specifications~\citep{freespec2, itree-server, itree-kv-server,
  robochart-itree}, embeddings~\cite{adam-binder,
  verify-effectful-haskell-in-coq, veriffi}, information-flow
analysis~\citep{itree-noninterference, dijkstra-forever},
testing~\citep{itree-testing},~\etc\@

However, \emph{the more expressive an interface is, the less we know about it}.

\paragraph{Motivating example}
To illustrate the problem, let's imagine that we are building an embedded
domain-specific language~(EDSL) for web services in Haskell. We can either get
from a server or post to a server in this EDSL.\@ The implementation of such get
and post operations can be different depending on the protocols and the web
service, so we want to have an abstract interface for these
operations.\footnote{The example is adapted from a similar example shown in
\citet[Section 1.3]{free-ap}.}

We can implement this interface using freer monads. There are many variants of
free and freer monad definitions, but, for simplicity, let's use the following
freer monads:
\begin{lstlisting}[language=haskell]
data FreerMonad (e :: Type -> Type) (a :: Type) where
  Ret :: a -> FreerMonad e a
  Eff :: e x -> (x -> FreerMonad e a) ->
         FreerMonad e a
\end{lstlisting}
A freer monad is an algebraic data type parameterized by an ``effect'' \cdh{e}
and a return type \cdh{a}. It has two data constructors: a \cdh{Ret} constructor
for pure computations; and an \cdh{Eff} constructor for effects and a
continuation that reacts to the result of an effect.

In our web service example, the effect can be defined as the following
generalized algebraic datatype~(GADT):
\begin{lstlisting}[language=haskell]
data WebServiceOps :: Type -> Type where
  Get  :: URL -> [String] -> WebServiceOps String
  Post :: URL -> [String] -> String -> WebServiceOps ()
\end{lstlisting}
Such an interface enables us to define the get and post operations as the
following functions:
\begin{lstlisting}[language=haskell]
get :: URL -> [String] ->
       FreerMonad WebServiceOps String
get url params = Eff (Get url params) Ret

post :: URL -> [String] -> String ->
       FreerMonad WebServiceOps String
post url params body = Eff (Post url params body) Ret
\end{lstlisting}
These functions simply ``embed'' an ``abstract syntax node'' representing a get
or a post operation in a freer monad. To implement the underlying operations, we
need to implement \emph{effect handlers}. For example, one possible effect
handler would have the following type signature:
\begin{lstlisting}[language=haskell]
handleWebServiceOps :: WebServiceOps a -> IO a
\end{lstlisting}
But the user can choose to implement such a handler using other monads as well.
For example, a user can implement a handler with a state monad to run
simulations without actually invoking I/O.

A key advantage of freer monads is that they can be interpreted into \emph{any
monads}, as long as there is an effect handler for such monad. More concretely,
we can define the following interpretation function for freer monads:
\begin{lstlisting}[language=haskell]
interp :: Monad m => (forall a. e a -> m a) ->
                     FreerMonad e a -> m a
interp _      (Ret a)     = return a
interp handle (Eff eff k) =
  handle eff >>= interp handle . k
\end{lstlisting}

\paragraph{Challenge: static analysis}
So far so good. But what if we want to do some static analysis on our EDSL?\@
Perhaps we want to count the number of get and post operations, respectively. Or
perhaps we want to collect on the URLs that we are posting to for security
checks. Since we are implementing an EDSL, we would like to implement such a
static analyzer \emph{as a function} inside Haskell as well.

Unfortunately, we cannot do that with this approach based on freer monads. This
is because monads are so expressive, that we can easily define the following
function:
\begin{lstlisting}[language=haskell]
postDepends :: URL -> [String] -> String ->
               FreerMonad WebServiceOps ()
postDepends url params body =
  get url params >>=
  postNTimes url params body .
  (read :: String -> Integer)
\end{lstlisting}
Where \cdh{postNTimes url params body n} posts to \cdh{url} with parameters
\cdh{params} and content body \cdh{body} for \cdh{n} times. In this
\cdh{postDepends} function, the number of post operations cannot be counted, as
the number depends on the result of the first get operation. We can only know
the result of a get operation \emph{after} we actually perform the get
operation.

However, it is unsatisfying that we cannot statically analyze the following
code, either:
\begin{lstlisting}[language=haskell]
echo :: URL -> URL -> [String] -> 
        FreerMonad WebServiceOps ()
echo url1 url2 params =
  get url1 params >>= post url2 params 
\end{lstlisting}
In this case, only the content body of the post operation depends on the result
of the get operation. If we look at the code, we know that there must be an
equal amount of get and post operations. Furthermore, if none of \cdh{url1} or
\cdh{url2} depend on an effectful operation, we should be able to collect all
the URLs we are getting from or posting to.

Sadly, we cannot write a function to statically analyze this piece of code,
either, due to the expressiveness of monads. Indeed, this dynamic nature of
monads has been discussed in literatures under various contexts~\citep{free-ap,
  adverbs, build, selective}.

\paragraph{Our work}
In this paper, we explore a different structure that offers an alternative
trade-off between expressiveness and static analyzeability: \emph{freer arrows}.
Arrows are an abstraction initially proposed by \citet{hughes_generalising_2000}
as a generalization of monads. Later, \citet{idioms-arrows-monads} discover that
arrows sit between applicative functors and monads in terms of expressiveness.
In this paper, we propose several types of \emph{freer arrows} and explore their
usefulness in Haskell.

Prior works like \citet{computation-monoids} have explored the concept of free
arrows, a datatype replying on \emph{profunctors}. Our definitions are
\emph{freer} because we remove the dependency on profunctors, following the
distinction set by \citet{freer}.

We make the following contributions:
\begin{itemize}
  \item We define three basic types of freer arrows, including \emph{freer
  pre-arrows}~(\cref{subsec:freer-pre}), \emph{freer
  arrows}~(\cref{subsec:freer-arrow}), and \emph{freer choice
  arrows}~(\cref{subsec:freer-choice}).
  \item We present a case study with freer arrows on \emph{choreographic
  programming}~(CP) based on the HasChor framework~\citep{haschor}. We show that
    are able to implement \emph{static endpoint projection}, \emph{endpoint
    static analysis}, and \emph{selective broadcasting} with HasChor implemented
    using freer arrows~(\cref{sec:case-state,sec:haschor-a}).
\end{itemize}

In addition, we provide an overview of our approach in \cref{sec:background}, we
discuss other aspects of freer arrows, related work, and future work in
\cref{sec:discussion,sec:related-work}. We conclude this paper in
\cref{sec:conclusion}.

%% file: background.tex
\section{Freer Arrows: An Overview}\label{sec:background}\label{sec:overview}

We show a slightly modified version of \citet{hughes_generalising_2000}'s arrows
in the top half of \cref{fig:arrow-core}. An arrow is first a \cdh{Category}. A
\cdh{Category} has two methods: \cdh{id} for identity category, and \cdh{(.)}
for composing two categories whose input and output match. We also define a
commonly used operator \cdh{(>>>)} for categories, which essentially flips the
arguments of \cdh{(.)}. A \cdh{PreArrow} adds to \cdh{Category} an additional
\cdh{arr} method that ``converts'' a function to a category.\@ An \cdh{Arrow}
adds an additional \cdh{first} method to a \cdh{PreArrow} that allows it to
carry extra values using a product in both its input and output.

\begin{figure}[t]
\begin{lstlisting}[language=Haskell]
class Category (cat :: k -> k -> Type) where
  id  :: forall (a :: k). cat a a
  (.) :: forall (b :: k) (c :: k) (a :: k).
         cat b c -> cat a b -> cat a c
         
-- A commonly used operator for categories
(>>>) :: forall k (a :: k) (b :: k) (c :: k)
         (cat :: k -> k -> Type).
  Category cat => cat a b -> cat b c -> cat a c
(>>>) = flip (.)
  
class Category a => PreArrow a where
  arr :: (b -> c) -> a b c

class PreArrow a => Arrow a where
  first :: a b c -> a (b, d) (c, d)

-- |- The state arrow.
newtype AState s a b =
  AState { runAState :: (a, s) -> (b, s) }

instance Category (AState s) where
  id = AState id
  AState f . AState g = AState $ f . g

instance PreArrow (AState s) where
  arr f = AState $ first f

instance Arrow (AState s) where
  first  (AState f) =
    AState $ \((a, c), s) ->
    let (b, s') = f (a, s) in ((b, c), s')
\end{lstlisting}
\caption{Key typeclass definitions related the arrows and
  state~(\ie~\cdh{AState}) as an instance of arrows. We defined \cdh{AState}
  with state \cdh{s}, input \cdh{a}, and output {b}. The function is
  intentionally uncurried, so we have simpler definition for methods like
  \cdh{id}, \cdh{(.)}, and \cdh{arr}. Note that we define these methods by using
  the \cdh{id}, \cdh{(.)}, and \cdh{first} methods of function arrows,
  respectively.}\label{fig:arrow-core}\label{fig:state-arrow}
\end{figure}

Compared with typeclasses like functors/applicative functors/monads, arrows
additionally contain the input type as part of its type.

Functions are a classic example of arrows. Every monad can also be made into an
arrow by adding the input to its type~\citep{hughes_generalising_2000}. For
example, state is an instance of arrows, as demonstrated in the bottom half of
\cref{fig:state-arrow}.

\paragraph{The freer arrow datatype}
The key idea of our approach is based on the freer arrow datatype. Given any
effect datatype \cdh{e} of kind \cdh{Type -> Type -> Type}, applying the freer
arrow datatype to \cdh{e} will result in an instance of \cdh{Arrow}. Here is one
definition of the freer arrow datatype~(or simply \emph{freer arrows} for short
in the rest of this paper):
\begin{lstlisting}[language=Haskell]
data FreerArrow e x y where
  Hom :: (x -> y) -> FreerArrow e x y
  Comp :: (x -> (a, c)) -> e a b ->
          FreerArrow e (b, c) y -> FreerArrow e x y
\end{lstlisting}
Freer arrows have two data constructors: \cdh{Hom} and \cdh{Comp}. The \cdh{Hom}
constructor ``embeds'' pure computations in a freer arrow. The \cdh{Comp}
constructor, on the other hand, ``embeds'' an effect \cdh{e} that has an input
\cdh{a} and output \cdh{b}. Together with the effect \cdh{e a b}, the \cdh{Comp}
constructor also contains (1)~a function for retrofiting an input \cdh{x} to the
effect's input \cdh{a} with a carried value of type \cdh{c}, and (2)~a
continuation freer arrow that takes the effect's output \cdh{b} as an input with
the previously carried value of type \cdh{c}. The carried value of type \cdh{c}
may look bizarre at this point, but it is crucial to \emph{strengthening} freer
arrows for implementing the \cdh{first} method of arrows. We defer the detailed
definitions of these methods to \cref{subsec:freer-pre,subsec:freer-arrow}.

\paragraph{The web service example, revisited}
To use freer arrows for our web service example, we can encode the web service
interface using the following effect datatype:
\begin{lstlisting}[language=haskell]
data WebServiceOps :: Type -> Type -> Type where
  Get  :: URL -> [String] -> WebServiceOps () String
  Post :: URL -> [String] -> WebServiceOps String ()  
\end{lstlisting}
Compared with the \cdh{WebServiceOps} shown in \cref{sec:intro}, this datatype
additionally contains the type of the input as its first type argument.

We can further define the following ``smart constructors'' for \cdh{Get} and
\cdh{Post}:
\begin{lstlisting}[language=haskell]
embed :: e x y -> FreerArrow e x y
embed f = Comp (,()) f (arr fst)
  
get :: URL -> [String] ->
       FreerArrow WebServiceOps () String
get url params = embed $ Get url params

post :: URL -> [String] ->
        FreerArrow WebServiceOps String ()
post url params = embed $ Post url params
\end{lstlisting}
The \cdh{embed} function ``embeds'' an effect in a \cdh{Comp}.\@ \cdh{Comp}
requires an existential type, \ie~the type of the carried value, but we do not
need to carry any values in this case, so we just use \cdh{()} as that type.
Then \cdh{get} and \cdh{put} are defined as \cdh{embed}ding \cdh{Get} and
\cdh{Put}, respectively.

We can now re-implement the \cdh{echo} function shown previously using freer
arrows:
\begin{lstlisting}[language=haskell]
echo :: URL -> URL -> [String] ->
        FreerArrow WebServiceOps () ()
echo url1 url2 params =
  get url1 params >>> post url2 params
\end{lstlisting}
Syntactically, this definition is almost the same as our previous definition of
\cdh{echo}, except that we replace \cdh{>>=} with \cdh{>>>}. However, because
this new function is defined using freer arrows, we \emph{can} do static
analysis on it now. For example, the following function counts the number of
effects in any programs implemented using freer arrows:
\begin{lstlisting}[language=Haskell]
count :: FreerArrow e x y -> Int
count (Hom _) = 0
count (Comp _ _ y) = 1 + count y
\end{lstlisting}

Just like freer monads, we can implement a freer arrow into any arrows, as long
as we can provide an effect handler:
\begin{lstlisting}[language=haskell]
interp :: Arrow arr =>
  (e :-> arr) -> FreerArrow e x y -> arr x y
interp _       (Hom f) = arr f
interp handler (Comp f x y) =
  arr f >>> first (handler x) >>> interp handler y
\end{lstlisting}
Where the type operator \cdh{:->} is defined as follows:\footnote{We borrow this
notation from the \cdh{profunctors} library by Edward Kmett:
\url{http://github.com/ekmett/profunctors/}.}
\begin{lstlisting}[language=haskell]
type p :-> q = forall a b. p a b -> q a b
\end{lstlisting}

\paragraph{Why static analysis matters}
It might seem that we are giving up the expressiveness of monads for the niche
advantage of counting the number of effects. However, the significance of being
able to define a function to statically analyze a program should not be
underestimated.

To demonstrate the usefulness of static analysis in pratice, we will show some
applications in \emph{choreographic programming}~(CP). In choreographic
programming, we write one program that will be \emph{projected} to multiple
\emph{endpoints} that are clients or servers participating in the
communications. \Citet{haschor} showed that CP can be implemented as an EDSL in
Haskell using freer monads as a library called HasChor.

In \cref{sec:haschor1,sec:haschor-a}, we re-implemented HasChor using freer
arrows. We are able to implement the following features that were not possible
in the original HasChor:
\begin{itemize}
  \item \emph{Static endpoint projection and endpoint static analysis}. We can
    perform \emph{endpoint projection} statically, so that each endpoint only
    has their ``full program'' before execution. These ``full programs''
    themselves are statically analyzeable, so we can inspect information such as
    if an endpoint is only communicating with trusted servers, or which
    endpoints are intensive on local computation and which endpoints heavy on
    network communications, \etc\@
  \item \emph{Selective broadcasting}. When a choice needs to be synchronized
    among multiple parties, we can statically analyze the choreography to find
    all the involved participants. In this way, we can project the choreography
    such that only involved parties will be targets of a broadcast. This allows
    a lightweight implementation of mini-\emph{conclaves}~\citep{conclave}.
\end{itemize}

\paragraph{What about conditionals or loops?}

But would we be limiting the expressiveness too much by using freer arrows? What
if we need conditionals or loops in our EDSL?\@

Implementing conditionals is straightforwad if we use a more expressive variant
of freer arrows: freer choice arrows. Freer choice arrows implement the
\cdh{ArrowChoice} typeclass that allows branching based on a sum type. We defer
more detailed discussion of \cdh{ArrowChoice} and freer choice arrows to
\cref{subsec:freer-choice}.

For loops, we consider two situations: when the number of iterations is
statically known, and when it's not. If the number of iterations is statically
known, we can implement such a loop by a recursion based on the number of
iterations. If the number of iterations is not statically known, we need to use
additional datatypes for \emph{reifying} loops combined with freer choice arrows
for modeling conditionals. We briefly discuss one such an example datatype in
\cref{sec:discussion,apx:elgot}. Notably, by using a datatype for loops, we lose
the ability to statically analyze the entire program. However, we are still able
to statically analyze the loop body or the part before/after the loop, if they
are defined using freer arrows.

\paragraph{What about other freer datatypes?}

Though we have demonstrated that freer arrows can help us with this web service
example, a reader might still (rightfully) wonder: Why do we choose freer
arrows? Why not use other freer datatypes like freer applicative
functors~\citep{free-ap} or freer selective functors~\citep{selective}? Existing
works have shown that we can define functions that perform static analysis on
these datatypes as well~\citep{free-ap, selective, adverbs}.

It turns out that arrows have just the right expressiveness for our purpose, a
fact that was first demonstrated by \citet{idioms-arrows-monads}. In particular,
they show that applicative functors are as expressive as any arrows \cdh{a} for
which there is an isomorphism between \cdh{a b c} and \cdh{a () (b -> c)}. This
isomorphism indicates that applicative functors are essentially arrows whose
input type can always be considered as \cdh{()}, \ie~an effectful computation
never uses the result of another effectful computation as input in applicative
functors. This means that we \emph{cannot} even implement the \cdh{echo}
function if we use a freer applicative functor!

%% file: freer-arrow.tex
\section{Freer Pre-Arrows}\label{subsec:freer-pre}

We show the definition of freer pre-arrows in \cref{fig:freer-prearrows}. A
freer pre-arrow is parameterized by an effect datatype \cdh{e} of kind \cdh{Type ->
Type -> Type}, an input datatype \cdh{x :: Type}, and an output datatype
\cdh{y :: Type}~(line 1). There are only two constructors in a freer pre-arrow.
The first constructor, \cdh{Hom}, simply wraps a function of type \cdh{x -> y}
inside it~(line 2). The second constructor, \cdh{Comp}, is the key for embedding
effects and composing freer arrows~(lines 3--5). There are two existential types
in the \cdh{Comp} constructor, namely \cdh{a} and \cdh{b}~(\ie~they don't appear
in the type of \cdh{FreerArrow}). The \cdh{Comp} constructor takes three
arguments. First, there is a function argument \cdh{x -> a} that does some pure
computation that transform an \cdh{x} to type \cdh{a}~(line 3). The value of
type \cdh{a} is then passed to the effect \cdh{e a b} that outputs a value of
type \cdh{b}~(line 3). Finally, there is another \cdh{FreerArrow} that takes the
value of type \cdh{b} and returns \cdh{y}~(line 4).

\begin{figure}[t]
\begin{lstlisting}[language=Haskell,numbers=left,xleftmargin=4.0ex]
data FreerPreArrow e x y where
  Hom   :: (x -> y) -> FreerPreArrow e x y
  Comp  :: (x -> a) -> e a b ->
           FreerPreArrow e b y ->
           FreerPreArrow e x y

-- | Freer pre-arrows are categories.
instance Category (FreerPreArrow e) where
  id = Hom id

  Hom f       . Hom g       = Hom (f . g)
  Comp f' e y . Hom g       = Comp (f' . g) e y
  f           . Comp f' e y = Comp f' e (f . y)

-- | Freer pre-arrows are pre-arrows.
instance PreArrow (FreerPreArrow e) where
  arr = Hom

-- | Embed an effect in freer arrows.                        
embed :: e x y -> FreerPreArrow e x y
embed f = Comp id f id

-- | The type for effect handlers.
type x :-> y = forall a b. x a b -> y a b

-- | Freer pre-arrows can be interpreted into any
-- pre-arrows, as long as we can provide an effect
-- handler.
interp :: Arrow arr =>
  (e :-> arr) -> FreerPreArrow e x y -> arr x y
interp _       (Hom f)       = arr f
interp handler (Comp f x y)  =
  arr f >>> (handler x) >>> interp handler y
\end{lstlisting}
\caption{Key definitions of freer pre-arrows~(\cdh{FreerPreArrow}) in Haskell.
  The code is simplified for presentation. In our artefact, we additionally show
  that \cdh{FreerPreArrow} is a \cdh{Profunctor}, we then use the \cdh{lmap}
  method of \cdh{Profunctor}s to define \cdh{(.)}. The definition shown here is
  equivalent to the definition based on
  \cdh{Profunctor}s.}\label{fig:freer-prearrows}
\end{figure}

To show that a freer pre-arrow is a \cdh{Category}, we need to define both the
\cdh{id} and \cdh{(.)} methods~(lines 7--13). The \cdh{id} method can be defined
as \cdh{Hom id}, where the \cdh{id} in the definition is the identity
function~(line 9). Composition \cdh{(.)} is defined by pattern matching on both
arguments. When both arguments are \cdh{Hom}s, the definition is just applying
\cdh{Hom} to the function composition~(line 11). When the first argument is
\cdh{Comp f' x y} and the second argument is \cdh{Hom g}, we simply compose
functions \cdh{f'} with \cdh{g}~(line 12). When the second argument is a
\cdh{Comp f' x y}, we make a recursive call to compose the first argument with
\cdh{y}~(line 13). Note that even though we use \cdh{(.)} in the definition in
every case, only the last one on line 13 is a recursive call; all other
\cdh{(.)} are function compositions.

Freer pre-arrows are also \cdh{PreArrow}~(lines 15--17). The \cdh{arr} method is
simply \cdh{Hom}~(line 17).

Given an effect \cdh{e x y}, we can \cdh{embed} the effect in a freer
pre-arrow~(lines 19--21). We embed such an effect using the \cdh{Comp}
constructor with an identity function~(the first \cdh{id} on line 21) and an
identity freer pre-arrow~(the second \cdh{id} on line 21).

Finally, we can interpret a \cdh{FreerPreArrow} to any pre-arrow if we provide
an ``effect handler''~(lines 23--33). An effect handler has type \cdh{e :-> arr}
where \cdh{e} is an effect type and \cdc{arr} is a pre-arrow. We use the type
operator \cdh{x :-> y} to represent ``transformations'' from \cdh{x a b} to
\cdh{y a b} for any input type \cdh{a} and output type \cdh{b}~(lines 23--24).
When interpreting a \cdc{FreerPreArrow}, we do a case analysis on the freer
pre-arrow. In the case of \cdh{Hom f}, we simply apply the \cdh{arr} method of
the pre-arrow \cdh{arr} to \cdh{f}~(line 31). In the case of \cdh{Comp f x y},
we use our handler to handle \cdh{x}, apply the \cdh{lmap} method of the
pre-arrow \cdh{arr} to both \cdh{f} and \cdh{handler x}, and compose it with the
recursive interpretation of \cdh{y}~(lines 32--33).

\paragraph{Expressiveness}
Freer pre-arrows are sufficient for defining functions like \cdh{echo} in our
web service example. Indeed, we can change the type signature of \cdh{get},
\cdh{post}, and \cdh{echo} functions defined in the previous section to use
\cdh{FreerPreArrow} instead.

However, freer pre-arrows only allows the use of effect in a streamlined manner.
For example, we cannot ``share'' the output of an effect as input of multiple
other effects. To do so, we need the \cdh{first} method defined on
\cdh{Arrow}s~(also known as the \emph{strength} of arrows). We talk about freer
arrows that allow this capability in \cref{subsec:freer-arrow}.

\paragraph{Static analysis}
We can statically analyze a freer pre-arrow to obtain its ``approximation''. For
example, the \cdh{count} function defined in the previous function also works on
freer pre-arrows. In fact, we can define the following generalized
\cdh{approximate} function:
\begin{lstlisting}[language=haskell]
approximate :: Monoid m =>
               (forall x y. e x y -> m) ->
               FreerPreArrow e a b -> m
approximate _ (Hom _) = mempty
approximate f (Comp _ e y) = f e <> approximate f y
\end{lstlisting}
Given an ``approximation function'' \cdh{f} that approximates an effect to a
monoid \cdh{m}, we can iterate over the entire freer pre-arrow and collect all
the effect approximations via \cdh{(<>)}.\footnote{The function does essentially
the same thing as \cdh{foldMap}, but it cannot be defined as a \cdh{foldMap} as
it has a different type signature.} For example, the \cdh{count} function is a
special case of the \cdh{approximate} function:
\begin{lstlisting}[language=haskell]
count :: FreerPreArrow e x y -> Int
count = getSum . approximate (const $ Sum 1)
\end{lstlisting}
Here we use the \cdh{Sum} type from Haskell's \cdh{Data.Monoid} module. Recall
that in Haskell, \cdh{Int}s are not \cdh{Monoid}s. We need to specify whether
the monoid operations correspond to plus or multiplication using the \cdh{Sum}
type or \cdh{Product} type.

We can do more than \cdh{count} with \cdh{approximate}. For example, we can also
collect a \emph{trace} of all the effects that will happen in a freer pre-arrow.
We defer more examples demonstrating usefulness of static analysis to
\cref{sec:haschor-a}.

\section{Freer arrows}\label{subsec:freer-arrow}

To enable ``sharing'' the result of an effect among multiple other effects
instead of only the next effect, we need to enhance freer pre-arrows with
\emph{products} to enable defining \cdh{first}. This gives us freer arrows. It
turns out that we only need to modify the \cdh{Comp} constructor to obtain freer
arrows. We show the definition in \cref{fig:freer-arrows}.

\begin{figure}[t]
\begin{lstlisting}[language=Haskell,numbers=left,xleftmargin=4.0ex]
data FreerArrow e x y where
  Hom :: (x -> y) -> FreerArrow e x y
  Comp :: (x -> (a, c)) -> e a b ->
          FreerArrow e (b, c) y ->
          FreerArrow e x y

-- | Freer arrows are arrows.
instance Arrow (FreerArrow e) where
  first :: FreerArrow e a b ->
           FreerArrow e (a, c) (b, c)
  first (Hom f) = Hom $ first f
  first (Comp f a b) =
    Comp (first f >>> assoc)
         a
         (lmap unassoc (first b))

-- | Embed an effect in freer arrows.                        
embed :: e x y -> FreerArrow e x y
embed f = Comp (,()) f (arr fst)

-- | Freer arrows can be interpreted into any
-- arrows, as long as we can provide an effect
-- handler.
interp :: Arrow arr =>
  (e :-> arr) -> FreerArrow e x y -> arr x y
interp _       (Hom f) = arr f
interp handler (Comp f x y) =
  arr f >>> (first (handler x)) >>>
  interp handler y

-- Helper functions. Definitions omitted.
assoc :: ((a,b),c) -> (a,(b,c))
unassoc :: (a,(b,c)) -> ((a,b),c)
\end{lstlisting}
\caption{Key definitions of freer arrows~(\cdh{FreerArrow}) in Haskell.
  Instances of \cdh{Category} and \cdh{PreArrow} are omitted as they are the
  same as those of \cdh{FreerPreArrow}s shown in
  \cref{fig:freer-prearrows}.}\label{fig:freer-arrows}
\end{figure}

The new \cdh{Comp} constructor contains three existential types: \cdh{a},
\cdh{b}, and \cdh{c}~(lines 3--4). The return type of the function argument is
changed to \cdh{(a, c)} to allow a carried value of type \cdh{c}~(line 3).
Correspondingly, the input type of the inner freer arrow is changed to \cdh{(b,
  c)}~(line 4). The effect type \cdh{e a b} remains unchanged. This means that
the value of type \cdh{c} is simply ``passed along'' to the next
\cdh{FreerArrow} without being processed by \cdh{e}---this may seem redundant
but it's crucial for implementing the \cdh{first} method~(also known as the
\emph{strength} of arrows).

\cdh{FreerArrow}s are instances of \cdh{Category}, and \cdh{PreArrow}. These
definitions are the same as those of freer pre-arrows, so we omit them here.

To show that \cdh{FreerArrow}s are an instance of \cdh{Arrow}, we need to define
the \cdh{first} method~(racall \cref{fig:arrow-core}). We show the type
signature of \cdh{first} for freer arrows in lines 9--10. We define this method
by pattern matching on \cdh{FreerArrow}s. In the case of \cdh{Hom f}, we apply
the \cdh{first} method of the function arrow to function \cdh{f}~(line 11). In
the case of \cdh{Comp f a b}, we need a few extra steps to take care of types.
First, we apply the \cdh{first} method of the function arrow to \cdh{f}, which
gives us a product type in the form of \cdh{((_, _), _)}~(line 13). To match the
type with that of the effect \cdh{a}, we then call \cdh{assoc} to convert the
product to the form of \cdh{(_, (_, _))}~(line 13). Finally, we call
\cdh{unassoc}, the inverse of \cdh{assoc} on the recursive evaluation \cdh{first
  b}~(line 15). Here, function calls to \cdh{assoc} and \cdh{unassoc} are only
to align types---they do not pose any computational significance, but we have to
carry them around when using \cdh{first} method on \cdh{FreerArrow}s.

Given an effect, we can \cdh{emded} it in a freer arrow using the \cdh{Comp}
constructor~(lines 17--19). The definition here is more complicated than that of
freer pre-arrows, because we need to apply \cdh{Comp} to a function argument and
a continuation arrow that contain a carried value. However, we do not need any
carried value at this point, so we can simply use \cdh{()}. Our function
argument wraps the input in a pair whose second element is \cdh{()}; our
continuation arrow uses \cdh{arr fst} to drop the \cdh{()} from the pair.

We can interpret a freer arrow to any arrows using the \cdh{interp}
function~(lines 21--29). Compared with the \cdh{interp} function of
\cdh{FreerPreArrow}, we need to call \cdh{first} on \cdh{handler x}, due to the
type change in \cdh{f}~(line 28). This is inevitable even for the segments of a
freer arrow that we do not use \cdh{first}.

\paragraph{Expressiveness}
Based on the \cdh{first} method, we can further define some other useful
combinators shown in \cref{fig:arrow-methods}. Thanks to these combinators, we
can now ``share'' the result of an effect thanks to these methods. Taking the
web service example again, we can define the following function that gets the
content from \cdh{url1}, and then forward it to both \cdh{url2} and \cdh{url3}:
\begin{lstlisting}[language=haskell]
forward :: URL -> URL -> URL -> [String] ->
           FreerArrow WebServiceOps () ()
forward url1 url2 url3 params =
  get url1 params >>>
  post url2 params &&& post url3 params >>>
  arr (const ())
\end{lstlisting}

\begin{figure}[t]
\ifextended
\begin{lstlisting}[language=haskell]
(***) :: Arrow a => a b c -> a d e -> a (b, d) (c, e)
f *** g = first f >>> arr s >>> first g >>> arr s
  where s (x,y) = (y,x)

(&&&) :: Arrow a => a b c -> a b c' -> a b (c,c')
f &&& g = arr (\b -> (b,b)) >>> f *** g

second :: Arrow a => a b c -> a (d,b) (d,c)
second = (id ***)
\end{lstlisting}
\else
\begin{lstlisting}[language=haskell]
(***) :: Arrow a => a b c -> a d e -> a (b, d) (c, e)
(&&&) :: Arrow a => a b c -> a b c' -> a b (c,c')
second :: Arrow a => a b c -> a (d,b) (d,c)
\end{lstlisting}
\fi
\caption{Other useful combinators of \cdh{Arrow} that can be defined using the
  \cdh{first} method~\citep{hughes_generalising_2000}. In Haskell, these
  functions are also defined as methods of \cdh{Arrow}s. An \cdh{Arrow} instance
  can be defined by either \cdh{first} or \cdh{(***).}}\label{fig:arrow-methods}
\end{figure}

Even though we now have the ability to ``pass down'' values, freer arrows are
\emph{very static}. We cannot write a program that has conditionally executed
branches. This means that every effect embedded in a freer arrow \emph{will}
happen, a property that may not always be desirable. Fortunately, we can further
enhance freer arrows to allow conditionals and branching. We will talk about
freer choice arrows, the datatype that allows these features in \cref{subsec:freer-choice}.

\paragraph{Static analysis}
Freer arrows can be statically analyzed in the same way as freer pre-arrows. In
fact, the \cdh{approximate} function defined in \cref{subsec:freer-pre} works
for freer arrows if we simply modify its type signature.

\paragraph{Why are freer arrows defined in this way?}
Our freer pre-arrow and freer arrow definitions are based on the free arrows of
\citet{computation-monoids}. However, \citeauthor{computation-monoids}'s
definition is based on another concept known as \emph{profunctors}. Their free
arrow is a pre-arrow only if the effect datatype \cdh{e} is a profunctor and it
is an arrow only if \cdh{e} is a \emph{strong profunctor}. Our definition
is \emph{freer}---a notation coined by \citet{freer}---because we do not have
such requirements on \cdh{e}. Our key observation is that we can inline a free
profunctor inside the free arrow of \citeauthor{computation-monoids}. After
that, we only need some simplifications to obtain \cdh{FreerPreArrow}s.
Similarly, we can inline free strong profunctors to obtain \cdh{FreerArrow}s. We
describe this process in more detail in
\cref{apx:free-to-freer}.

Another way of defining freer pre-arrows and freer arrows is \emph{reifying} all
the operators of pre-arrows and arrows, similar to the reified datatypes studied
by \citet{adverbs}. However, such an approach would result in a datatype with
more constructors, which brings additional trouble to static analysis because we
need to match all constructors. Our definition, in comparison, represents
a \emph{normalized} freer arrow.

Finally, we can define freer pre-arrows and freer arrows using final
encodings, \ie~defining them by their interpreters. For example, freer arrows
can be defined as follows:
\begin{lstlisting}[language=haskell]
newtype FreerArrowFinal e b c = FreerArrowFinal {
  runFreer :: forall a. Arrow a =>
              (e :-> a) -> a b c }
\end{lstlisting}
We choose not to use this version because it does not have any data
constructors, which makes it difficult to perform any static analysis that
cannot be defined by \cdh{runFreer}. If we want, we can always interpret our
freer arrows to this \cdh{FreerArrowFinal}.

%% file: freer-choice.tex
\section{Freer Choice Arrows}\label{subsec:freer-choice}

When conditionals and branching are required, we need freer choice arrows. We
show the key definitions of freer choice arrows in
\cref{fig:freer-choice-instances}.

\begin{figure}[t]
\begin{lstlisting}[language=Haskell,numbers=left,xleftmargin=4.0ex]
data FreerChoiceArrow e x y where
  Hom :: (x -> y) -> FreerChoiceArrow e x y
  Comp :: (x -> Either (a, c) w) ->
          e a b ->
          FreerChoiceArrow e (Either (b, c) w) y ->
          FreerChoiceArrow e x y

instance Arrow (FreerChoiceArrow e) where
  first (Hom f) = Hom $ first f
  first (Comp f a b) =
    Comp (first f >>> distr >>> left assoc)
         a
         (lmap (left unassoc >>> undistr)
               (first b))

-- | The ArrowChoice typeclass.         
class Arrow a => ArrowChoice a where
  left :: a b c -> a (Either b d) (Either c d)

-- | Freer choice arrows are an instance of
-- [ArrowChoice].
instance ArrowChoice (FreerChoiceArrow e) where
  left (Hom f) = Hom $ left f
  left (Comp f a b) =
    Comp (left f >>> assocsum)
         a
         (lmap unassocsum (left b))

-- Helpful functions. Definitions omitted.
distr   :: (Either (a, b) c, d) ->
           Either ((a, b), d) (c, d)
undistr :: Either ((a, b), d) (c, d) -> 
           (Either (a, b) c, d)
assocsum   :: Either (Either x y) z ->
              Either x (Either y z) 
unassocsum :: Either x (Either y z) ->
              Either (Either x y) z
\end{lstlisting}  
\caption{Key definitions for \cdh{FreerChoiceArrow}. Instances of
  \cdh{Profunctor}, \cdh{Category}, and \cdh{PreArrow} are omitted as they are
  the same as those of \cdh{FreerPreArrow}s shown in
  \cref{fig:freer-prearrows}.}\label{fig:freer-choice-instances}
\end{figure}

The \cdh{Hom} constructor of \cdh{FreerChoiceArrow}~(line 2) is the same as that
of \cdh{FreerArrow}. However, the \cdh{Comp} constructor adds one additional
existential variable \cdh{w}~(lines 3--6). The function argument can either
return a product of type \cdh{(a, c)}, just like in \cdh{FreerArrow}, or return
a value of type \cdh{w}~(line 3). In the first case, the value of type \cdh{a}
will be passed to the effect of type \cdh{e a b} while the value of type \cdh{c}
is passed along~(line 4). In the second case, the value of type \cdh{w} will be
passed along without invoking the effect---representing a branch that the effect
does not happen. Finally, the inner \cdh{FreerChoiceArrow} needs to handle the
new input type \cdh{Either (b, c) w}~(line 5).

We also show the definitions of \cdh{first} method in lines 8--14. In the
\cdh{Hom} case, we simply use the \cdh{first} method of function arrows~(line
9).
\ifextended
The \cdh{Comp} case is more complex. We first apply the \cdh{first} method of
function arrows to \cdh{f}, which gives us a type of the form \cdh{((Either (_,
  _) _), _)}. We use the \cdh{distr} function defined on line 27 to convert the
type to a form of \cdh{Either ((_, _), _) (_, _)}.\footnote{This corresponds to
the distributivity of multiplication over additions, hence the function name
\cdh{distr}.} After that, we apply the \cdh{left} method of function choice
arrows to \cdh{assoc} to convert the type from the previous form to \cdh{Either
  (_, (_, _)) (_, _)}~(line 10). Finally, we apply \cdh{left unassoc >>>
  undistr}, which is the inverse of \cdh{distr >>> left assoc}, to the recursive
call~(line 11). Note that \cdh{undistr}, the inverse of \cdh{distr}, is defined
on line 28.
\else
The \cdh{Comp} case is more complex, but the key idea is using helper functions
shown in lines 30--37 to shuffle around the values to match the types. These
functions do not pose computational significance.
\fi

In addition, we show that freer choice arrows are an instance of
\cdh{ArrowChoice}.\@ We show the definition of \cdh{ArrowChoice} in lines
16--18. The key method of \cdh{ArrowChoice} is \cdh{left}, which branches on a
sum type \cdh{Either b d}. In the left case, it runs the arrow \cdh{a b c} and
returns as a left injection of \cdh{Either c d}. In the right case, it does
nothing and simply returns the value of type \cdh{d} as a right injection of
\cdh{Either c d}. The definition of \cdh{left} method on freer choice arrow is
shown in lines 20--27. Like the \cdh{first} method, most of the code on lines
25--27 is to shuffle around values to match the types.

\begin{figure}[t]
\begin{lstlisting}[language=haskell]
embed :: e x y -> FreerChoiceArrow e x y
embed f = Comp (Left . (,())) f (arr (fst ||| id))
    
interp :: (Profunctor arr, ArrowChoice arr) =>
          (e :-> arr) ->
          FreerChoiceArrow e x y -> arr x y
interp _       (Hom f) = arr f
interp handler (Comp f x y) =
  lmap f (left (first (handler x))) >>>
  interp handler y
\end{lstlisting}
\caption{The \cdh{embed} and \cdh{interp} methods of freer choice
  arrows.}\label{fig:freer-choice-embed-interp}
\end{figure}

We define the \cdh{embed} and \cdh{interp} functions of freer choice arrows in
\cref{fig:freer-choice-embed-interp}. The \cdh{embed} function wraps an effect
inside the \cdh{Comp} constructor. Given any input, we first combine it with
\cdh{()} to make a pair, then make the pair a left injection of a sum type. This
is encoded by the function argument \cdh{Left . (,())}. After executing the
effect, we pattern match on the result of the arrow. If we are on the left
branch, we discard the second element in the pair; if we are on the right
branch, we do nothing. This is encoded by the continuation arrow \cdh{arr (fst
  ||| id)}. We will explain what \cdh{(|||)} operator means shortly in this
section.

The \cdh{interp} function is similar to that of freer arrows, except that, in
the \cdh{Comp} case, we need to apply the \cdh{left} method after applying
\cdh{first} to the result of \cdh{handler} application.

\begin{figure}[t]
\ifextended
\begin{lstlisting}[language=haskell]
(+++) :: ArrowChoice a =>
         a b c -> a b' c' ->
         a (Either b b') (Either c c')
f +++ g = left f >>> arr mirror >>>
          left g >>> arr mirror
  where
    mirror :: Either x y -> Either y x
    mirror (Left x) = Right x
    mirror (Right y) = Left y

(|||) :: ArrowChoice a =>
         a b d -> a c d -> a (Either b c) d
f ||| g = f +++ g >>> arr untag
  where
    untag (Left x) = x
    untag (Right y) = y

right :: ArrowChoice a =>
         a b c -> a (Either d b) (Either d c)
right = (id +++)
\end{lstlisting}
\else
\begin{lstlisting}[language=haskell]
(+++) :: ArrowChoice a =>
         a b c -> a b' c' ->
         a (Either b b') (Either c c')
(|||) :: ArrowChoice a =>
         a b d -> a c d -> a (Either b c) d
right :: ArrowChoice a =>
         a b c -> a (Either d b) (Either d c)
\end{lstlisting}
\fi
\caption{Other useful combinators of \cdh{ArrowChoice} that can be defined using
  the \cdh{left} method~\citep{hughes_generalising_2000}. In Haskell, these
  functions are also defined as methods of \cdh{ArrowChoice}s. An
  \cdh{ArrowChoice} instance can be defined by either \cdh{left} or
  \cdh{(+++).}}\label{fig:arrowchoice-methods}
\end{figure}

\paragraph{Expressiveness}
Based on the \cdh{left} method, we can further define some other useful
combinators shown in \cref{fig:arrowchoice-methods}. If we replace the freer
arrows used in the web service example with freer choice arrows, we can then
implement programs that send messages to different \cdh{URL}s depending on a
result. For example:
\begin{lstlisting}[language=haskell]
forwardIf :: URL -> URL -> URL -> [String] ->
             String -> String ->
             FreerChoiceArrow WebServiceOps () ()
forwardIf url1 url2 url3 params m1 m2 =
  get url1 params >>> arr (read :: String -> Int) >>>
  arr (\n -> if n > 0 then Left m1 else Right m2) >>>
  post url2 params ||| post url3 params
\end{lstlisting}
Here, we replace all the \cdh{get} and \cdh{post} functions with the version
using freer choice arrows. The program first \cdh{get}s a message from
\cdh{url1} and parses it as an \cdh{Int}. If the number if greater than 0, we
\cdh{post} to \cdh{url2} with message \cdh{m1}; otherwise, we \cdh{post} to
\cdh{url3} with message \cdh{m2}.

Note that freer choice arrows only enable \emph{finite} branching. We still
cannot define functions like \cdh{postDepends} shown in \cref{sec:intro}, as
that function requires \emph{infinite} branching, \ie~branching on all unbounded
integers.

\ifextended
\paragraph{Static analysis}
Branching in freer choice arrows allows us to skip any effects. This means that
we can no longer accurately count the number of effects in a freer choice arrow.
However, we can still approximate a freer choice arrow, except that such an
approximation is an \emph{over-approximation}, reflected by the following
\cdh{overApprox} function:
\begin{lstlisting}[language=Haskell]
overApprox :: Monoid m =>
              (forall x y. e x y -> m) ->
              FreerChoiceArrow e a b -> m
overApprox _ (Hom _) = mempty
overApprox f (Comp _ e y) = f e <> overApprox f y
\end{lstlisting}
The function is in fact the same as the \cdh{approximate} function shown in
\cref{subsec:freer-pre}. This is no coincidence, as we calculate the
over-approximation by assuming no effect will be skipped.

We can also get an under-approximation of any freer choice arrows, but that
would not be interesting as the under-approximation will always be 0, \ie~we
skip all the effects.
\else
\paragraph{Static analysis}
Branching in freer choice arrows allows us to skip any effects, so we can no
longer accurately count the number of effects in a freer choice arrow. However,
we can still approximate a freer choice arrow, except that such an approximation
is an \emph{over-approximation}. Such a function is in fact the same as the
\cdh{approximate} function shown in \cref{subsec:freer-pre}. This is no
coincidence, as we calculate the over-approximation by assuming all effects will
happen.

We can also get an under-approximation of any freer choice arrows, but that
would not be interesting as the under-approximation will always be 0, \ie~we
skip all the effects.
\fi

%% file: haschor.tex
\section{Case Study: HasChor}\label{sec:case-state}\label{sec:haschor1}\label{sec:haschor-a}

In this section, we demonstrate the usefulness of freer arrows via a case study
on HasChor, an EDSL for \emph{choreographic programming}~(CP) in
Haskell~\citep{haschor}.

CP is a programming paradigm where one writes a single program, a
\emph{choreography}, that describes the entire behavior of a distributed system.
A separate process called \emph{endpoint projection} compiles a choreography to
correct-by-construction individual programs for each node.

The key idea of HasChor is that we can represent the program as a freer monad
that encodes the communications among multiple parties. Thanks to freer monads,
a user can re-use all existing advanced features from Haskell in HasChor,
including higher-order functions, type systems,~\etc\@ For example, we can write
a choreography for our previous \cdh{echo} example as follows:
\begin{lstlisting}[language=haskell]
echo :: Choreo IO (String @ "client")
echo = do
  strAtClient <- client `locally` (const getInput)
  strAtServer <- (client, strAtClient) ~> server
  (server, strAtServer) ~> client
\end{lstlisting}
The \cdh{echo} function has type \cdh{Choreo IO (String @ "client")}. It means
that it is a choreography that uses \cdh{IO} for all local computations and
returns a \cdh{String} at a location known as \cdh{"client"}. The type operation
\cdh{a @ l} is a customized definition in HasChor that represent a value of type
\cdh{a} at location \cdh{l}.

In this choreography \cdh{echo}, the client first tries to get an input string
from users \cdh{locally} and bind it to \cdh{strAtClient}. It then sends the
string to the server via the \cdh{(\~>)} operator. The server receives the
string as \cdh{strAtServer}. Finally, the server echoes the string back to the
client using the \cdh{(\~>)} operator again. The \cdh{locally} and \cdh{(\~>)}
operators have the following types:
\begin{lstlisting}[language=haskell]
locally :: Proxy l -> (Unwrap l -> m a) -> 
           Choreo m (a @ l)
(~>)    :: (Proxy l, a @ l) -> Proxy l' ->
           Choreo m (a @ l')
\end{lstlisting}
Here we use Haskell's \emph{phantom parameter}
\cdh{Proxy}\footnote{\url{https://hackage.haskell.org/package/base/docs/Data-Proxy.html}}
to inform Haskell's type checker which location we are at. The \cdh{locally}
function unwraps a local value at location \cdh{l} and executes it in a local
monad \cdh{m}~(\cdh{IO} in the case of \cdh{echo}). The \cdh{(\~>)} operator
sends a value of type \cdh{a} at location \cdh{l} to another location \cdh{l'}.
This results in choreography with that value at location \cdh{l'}.

Note that how this choreography ``takes no side''. It describes what happens
from a global view.

We can then \emph{project} this choreography to each node---in this case, the
client and the server---by interpreting all the multi-party communication events
to \emph{events specific to this endpoint}. This step is known as the endpoint
projection. On the server side, we get a similar program to the \cdh{echo}
function described in \cref{sec:intro}. On the client's side, we get a program
that first tries to get an input from the user, then sends it to the server, and
finally waiting to receive the echo from the server.

How is the endpoint projection implemented? Well, by interpreting freer monads!
Indeed, the definition of endpoint projection in HasChor is essentially as
follows:
\begin{lstlisting}[language=haskell]
epp :: Choreo m b a -> LocTm -> Network m b a
epp c l = interp handle c
  where handle = ... -- Definition omitted.
\end{lstlisting}
Both \cdh{Choreo} and \cdh{Network} in the type signature are freer monads. We
show the definitions of \cdh{Choreo} and \cdh{Network} in
\cref{fig:haschor-choreo-monads}. The \cdh{Choreo} type contains two effect
events: \cdh{Local}, which represents local computation at a location, and
\cdh{Comm}, which represents a communication between two locations. The
\cdh{Network} type contains three effect events: \cdh{Run}, a local computation;
\cdh{Send}, which sends a message to another location; and \cdh{Recv}, which
waits to receive a message from another location. For now, let's ignore
\cdh{Cond} in \cdh{ChoreoSig} and \cdh{BCast} in \cdh{NetworkSig}. The local
function \cdh{handle} ``translates'' events in \cdh{ChoreoSig} to events in
\cdh{NetworkSig}. Such ``translation'' depends on \cdh{l}, the second argument
of \cdh{epp}, which represents which endpoint we are currently projecting to.
For example, projecting \cdh{(client, strAtClient) \~> server} on the
\cdh{client} results in a \cdh{Send} event; but projecting it on the
\cdh{server} results in a \cdh{Recv} event.

\paragraph{Dynamic \vs static endpoint projection}
One issue with HasChor is that the endpoint projection is too \emph{dynamic} due
to freer monads. To understand why, let's revisit the \cdh{interp} function of
freer monads:
\begin{lstlisting}[language=haskell]
interp :: Monad m => (forall a. e a -> m a) ->
                     FreerMonad e a -> m a
interp _      (Ret a)     = return a
interp handle (Eff eff k) =
  handle eff >>= interp handle . k
\end{lstlisting}
The function pattern matches on the \cdh{FreerMonad} and applies \cdh{handle}
when it's an \cdh{Eff} case. However, only the first \cdh{interp} is fully
applied in this function. All recursive calls to \cdh{interp} are only partially
applied! In fact, these recursive calls cannot be fully applied until we receive
a result from the previous \cdc{handle eff}, and we can only get a result from a
\cdh{handle eff} when we actually execute the program on a node. This means what
each endpoint carries is not just their own program, but a ``suspended''
interpretation of the entire choreography. This is a known issue of HasChor. For
example, \citet{partition} try to solve this issue by using Template Haskell and
compiler plugins.

\begin{figure}[t]
\begin{lstlisting}[language=haskell]
type Choreo m = FreerMonad (ChoreoSig m)

data ChoreoSig m a where
  Local :: KnownSymbol l => Proxy l ->
           (Unwrap l -> m a) ->
           ChoreoSig m (a @ l)
  Comm :: (Show a, Read a,
           KnownSymbol l, KnownSymbol l') =>
          Proxy l -> a @ l ->
          Proxy l' -> ChoreoSig m (a @ l')
  Cond :: (Show a, Read a, KnownSymbol l) =>
          Proxy l -> a @ l ->
          (a -> Choreo m b) -> ChoreoSig m b

type Network m = FreerMonad (NetworkSig m)

data NetworkSig m a where
  Run :: m a -> NetworkSig m a
  Send :: Show a => a -> LocTm -> NetworkSig m ()
  Recv :: Read a => LocTm -> NetworkSig m a
  BCast :: Show a => a -> NetworkSig m ()
\end{lstlisting}
\caption{The \cdh{Choreo} type and \cdh{Network} type in HasChor implemented
  using freer monads.}\label{fig:haschor-choreo-monads}
\end{figure}

We can solve this issue without metaprogramming by replacing freer monads with
freer arrows. The change involves changing a number of types and replacing
monadic operations with arrow operations, but most of these changes are
straightforward. We show the new type definitions of \cdh{Choreo} and
\cdh{Network} in \cref{fig:choreo-and-network}. Interested readers can find a
detailed implementation of our work in our supplementary materials, which will
be released as a publicly-available artifact.

With freer arrows, we can re-implement the previous \cdh{echo} choreography as
follows:
\begin{lstlisting}[language=haskell]
echo = client `locally0` getInput >>>
       client ~> server >>>
       server ~> client
\end{lstlisting}

The new \cdh{epp} function also uses the \cdh{interp} method of freer
arrows~(\cref{fig:freer-arrows}) instead. Unlike the \cdh{interp} method of
freer monads, recursive calls in \cdh{interp} of freer arrows are always fully
applied.

\paragraph{Endpoint static analysis}
What do we gain by having a static endpoint projection? For a start, we can now
statically analyze every endpoint. For example, the following function collects
all the effect events happened at an endpoint:
\begin{lstlisting}[language=haskell]
collect :: Network ar a b -> [Event]
collect = approximate (singleton . trace)
  where trace :: NetworkSig ar a b -> Event
        trace = ... -- Definition omitted.
\end{lstlisting}
The function works on the \cdh{Network} type and collects a list of effectful
events of \cdh{Local}, \cdh{Send}, and \cdh{Recv}. Thanks to the static nature
of freer arrows, we can fully apply the \cdh{epp} function to a program of
\cdh{Choreo} type to obtain a program of \cdh{Network} type at an endpoint. We
can then apply this \cdh{collect} function to statically analyze the program at
an endpoint \cdh{l}.

This can be useful for several reason. For example, we can check what other
endpoints that \cdh{l} communicates with. If \cdh{l} is a secret endpoint that
only communicates with other trusted endpoints, we can use this static analysis
to ensure that all the endpoints \cdh{l} talk to are trusted. We can also check
which endpoints are intensive on local computations and which endpoints are
heavy on network communications. We can then use this information to for better
resource allocation or endpoint placement.

\begin{figure}[t]
\begin{lstlisting}[language=Haskell]
type Choreo ar = FreerArrow (ChoreoSig ar)

data ChoreoSig ar b a where
  Local :: KnownSymbol l => Proxy l -> ar b a ->
           ChoreoSig ar (b @ l) (a @ l)
  Comm :: (Show a, Read a,
           KnownSymbol l, KnownSymbol l') =>
          Proxy l -> Proxy l' ->
          ChoreoSig ar (a @ l) (a @ l')
  Cond :: (Show b, Read b, KnownSymbol l) =>
          Proxy l -> Choreo ar b a ->
          ChoreoSig ar (b @ l) a

type Network ar = FreerArrow (NetworkSig ar)

data NetworkSig ar b a where
  Run :: ar b a -> NetworkSig ar b a
  Send :: Show a => LocTm -> NetworkSig ar a ()
  Recv :: Read a => LocTm -> NetworkSig ar () a
  -- [BCast] includes an extra argument to support
  -- selective broadcasting.
  BCast :: Show a => HashSet LocTm ->
           NetworkSig ar (a @ l) a
\end{lstlisting}
\caption{The \cdh{Choreo} type and \cdh{Network} type in HasChor implemented
  using freer arrows.}\label{fig:choreo-and-network}
\end{figure}


\paragraph{Selective broadcasting}
One feature in HasChor that we have not discussed so far is \cdh{cond}, a
construct that implements the \emph{knowledge of
choice}~\citep{carbone-montesi-deadlock-freedom-by-design,hirsch-garg-pirouette,choral}.

We show an example choreography of a key-value server that uses \cdh{cond} in
\cref{fig:kvs-haschor}. The choreography contains a client, a primary server,
and a backup server. The client first prepares and sends a request to the
primary server. The primary server then handles the request locally. It then
branches depending on whether the request is a \cdh{Get} or a \cdh{Put}. In the
case of a \cdh{Get}, nothing special needs to be done. In the case of a
\cdh{Put}, however, the primary server first forwards the request to the backup
server, so that the backup server handles the request as well. Finally, no
matter the result is a \cdh{Get} or a \cdh{Put}, the primary responds to the
client with the result.

\begin{figure}[t]
\ifextended
\begin{lstlisting}[language=haskell]
kvs :: Choreo IO (Response @ "client")
kvs = do
  -- client prepares and sends the request
  req <- client `locally` getRequest
  req' <- (client, req) ~> primary

  -- primary handles the request
  res <- primary `locally` handleRequest req'
  -- disparate behavior based on the request
  cond (primary, req') \case
    Get k -> return ()
    Put k v -> do
      -- primary propagates and backup handles the
      -- request
      req'' <- (primary, req') ~> backup
      backup `locally` handleRequest req''
      return ()

  (primary, res) ~> client
\end{lstlisting}
\else
\begin{lstlisting}[language=haskell]
kvs :: Choreo IO (Response @ "client")
kvs = do
  req  <- client `locally` getRequest
  req' <- (client, req) ~> primary

  res  <- primary `locally` handleRequest req'
  cond (primary, req') \case
    Get k -> return ()
    Put k v -> do
      req'' <- (primary, req') ~> backup
      backup `locally` handleRequest req''
      return ()

  (primary, res) ~> client
\end{lstlisting}
\fi
\caption{An example key-value server with a client, a primary server, and a
  backup server implemented in HasChor.}\label{fig:kvs-haschor}
\end{figure}

The \cdh{cond} function is a thin wrapper of the \cdh{Cond} constructor in
\cdh{ChoreoSig} in HasChor~(\cref{fig:haschor-choreo-monads}). It takes a
location \cdh{l}, a value at location \cdh{l}, and a continuation based on the
value.

How do we deal with \cdh{Cond} in an endpoint projection? Because there might be
additional participants involved in the continuation of \cdh{Cond}, \eg~the
backup server is involved in our \cdh{kvs} choreography, so we need to make sure
all these involved participants synchronize, \ie~they all enter the same
conditional branch.

In HasChor, such a synchronization is implemented as a \emph{broadcast}. Taking
\cdh{kvs} as an example, the primary server will broadcast the request
\cdh{req'} to all other endpoints. In endpoint projection, this means that
\cdh{Cond} will be ``translated'' to the \cdh{BCast} event of
\cdh{NetworkSig}~(\cref{fig:haschor-choreo-monads}) to represent a broadcast. In
the meantime, all other endpoints will try to \cdh{Recv} this message, so that
they can enter the right branch depending on the same \cdh{req'}.

Ideally, this synchronization message should only be broadcast to endpoints that
are involved. Unfortunately, finding out all involved parties requires a static
analysis, which cannot be performed on freer monads! For this reason, HasChor
chooses to broadcast the synchronization message to \emph{all endpoints}
instead. In the \cdh{kvs} example, this means that \cdh{client} is forced to
participate in the synchronization, even though it is never mentioned in
\cdh{cond}.

Such implementation is inefficient and scales poorly with more nodes. One recent
work that tackles this issue is conclaves~\citep{conclave}, which relies on
dependent types.

However, \emph{freer choice arrows} enable a more lightweight approach that
does not require more complicated type systems. We show the \cdh{Cond}
constructor of \cdh{ChoreoSig} in our version with freer arrows in
\cref{fig:choreo-and-network}.\footnote{Freer choice arrows are not required for
implementing constructors like \cdh{Cond}, but all the interesting cases with
\cdh{Cond} only happen with branching~(\eg~the branching based on requests in
\cdh{kvs}), so we need freer choice arrows to implement the case study.} Note
that the continuation inside \cdh{Cond} is simply represented by a freer choice
arrow, so it can be statically analyzed. Indeed, in our implementation of
\cdh{epp}, we first use a static analysis to gather all the participants in this
``continuation choreography''. After that, we interpret \cdh{Cond} to a modified
\cdh{BCast} event~(\cref{fig:choreo-and-network}) that carries an additional
argument representing a set of locations. In this way, we only \cdh{BCast} to
locations that are known to be involved due to static analysis. We show our
implementation of \cdh{kvs} using freer arrows in \cref{apx:kvs-arrows}.

%% file: discussion.tex
\section{Discussion and Future Work}\label{sec:discussion}

\paragraph{Composing effects}
One advantage of freer monads or their variants is that we can compose effects
to implement \emph{algebraic effects}. Composing effects is fair game to freer
arrows as well. We show an example of implementing effect composition in
\cref{apx:composable-eff}.

\paragraph{Composing freer arrows}
An alternative approach to building a hierarchy of freer arrows is
\emph{composable} freer arrows. For example, we can have all three variants of
\cdh{Comp} constructors in freer pre-arrows, freer arrows, and freer choice
arrows modeled separately, then allow users to compose them as needed. However,
since the \cdh{Comp} constructors are recursive, we will run into the
\emph{expression problem}~\citep{expression}. One way to work around this is
using Church encodings~\citep{dtc,mtc}. Indeed, \citet{adverbs} have
successfully applied this approach in the context of freer functors/applicative
functors/monads. Overcoming the expression problem is out of the scope of this
work, but interested readers can find more recent exploration on this topic in
\citet{persimmon, compositional-programming, bowtie},~\etc\@

\paragraph{Loops}
With freer arrows, we can only implement loops that have a fixed number of
iterations. To implement other types of iterations, we need additional datatypes
that reify those iterations. We show an example datatype that reifies the Elgot
iteration~\citep{elgot, elgot-theories} in \cref{apx:elgot}.

\paragraph{Optimizations}
Static analysis usually helps with optimizations. In the future, we would like
to look into optimizations based on freer arrows. A challenge is that the
function arguments in \cdh{Comp} constructors are not amenable to static
analysis. One way to deal with it is \emph{defunctionalizing} common functions
used for composing freer arrows, similar to \citet{frp_restated}. We have a
prototype for these ``defunctionalized'' arrows but more experiments are
required to implement useful optimizations.

\paragraph{Metaprogramming}
In the future, we would like to explore combining freer arrows with
metaprogramming. Since freer arrows enable static analysis, we should be able to
exploit that to generate efficient code at compile time. Indeed, existing
approaches have explored the use of applicative functors in
metaprogramming~\citep{selective-parser, phases}. We believe arrows should open
up more opportunities. One challenge is again the function arguments in
\cdh{Comp} constructors---we cannot simply ``lift'' them to quoted code in
Template Haskell, but we might be able to do that if we limit the flexiblility
of functions in arrows.

%% file: related-work.tex
\section{Related Work}\label{sec:related-work}\label{sec:freers}

\paragraph{Arrows} \citet{hughes_generalising_2000} initially proposes arrows as
a generalization of monads. However, the relation among monads, applicative
functors~(also known as idioms at that time)~\citep{applicative}, and arrows
were unknown at that time. It was only later shown by
\citet{idioms-arrows-monads} that arrows are between applicative functors and
monads in terms of the expressiveness. Similar to monads, a \cdh{do}-notation
for arrows can also be used in writing arrows, instead of using primitive
methods like \cdh{first} and \cdh{left}~\citep{paterson_new_2001,
  paterson_arrows_2003}.

Arrows have been commonly used in domains like functional reactive
programming~\citep{frp_rearranged, frp_restated,frp-refactored,
  hudak_arrows_2003}. For example, \citet{hudak_arrows_2003} show that arrows
are useful for encoding behaviors in robots that combine continuous and discrete
parts, such as integration or derivation of sensor signals and changing between
finite program states.

In other domains, \citet{carette_how_2024} use arrows to represent quantum
computations as classical computations. Notably, in this computation model, the
underlying classical language is restricted to reversible functions, unlike the
unrestricted pure functions in the Haskell version. This brings up an
interesting question about one of the defining arrow operators \cdh{arr}, which
normally lifts a ``pure'' function into the arrow. We leave it to future work to
consider what notions of pure functions can be lifted into an arrow, and whether
it generalizes to, for example, any underlying category or profunctor.

\paragraph{Free structures} We discussed freer monads and their usefulness in
\cref{sec:intro}. Other freer structures have also been studied. For example,
\citet{free-ap} propose a version of freer applicative functors and show that
they can be statically analyzed. \Citet{selective} propose freer selective
applicative functors, which have been used by \citet{selective-parser} to
implement efficient staged parser combinators. \Citet{computation-monoids} study
various free structures, including free monads, free applicative functors, and
free arrows. By applying the techniques used by \citet{freer} to derive freer
monads, we can derive freer versions of these structures---in fact, we initially
derive our versions of freer arrows in this way.

Free structures enable a mixed embedding that has both semantics parts and
syntactic parts in the same data structure. This use case of free structures has
been studied by various works~\citep{adam-binder, adverbs, veriffi,
  folding-dsl}.

\paragraph{Algebraic effects with arrows} 
More recently, \citet{sanada_algebraic_2024} describes a semantics for writing
and interpreting effect handlers using arrows. This is parallel to our work,
because it involves defining a language with a structure for writing
user-defined effect handlers in that language. Our work does not involve a
bespoke language with effects, but rather describes a definition within an
existing language (Haskell) for freer arrows, along with tooling to use it for
extensible effects.

\paragraph{Choreographic Programming}
The earliest incarnation of choreographies can be traced back to the mid-2000s
in the context of web services~\citep{w3c-choreography-model, w3c-cdl,
  w3c-cdl-primer}. At the time, choreographies were mainly used as
\emph{specifications} of communicating processes.
\Citet{carbone-montesi-deadlock-freedom-by-design, montesi-dissertation}
pioneered choreographic \emph{programming} languages with their Chor language
and provided a correctness proof in terms of multiparty session
types~\citep{honda-mpsts}. Since then, much work has focused on developing the
theoretical foundations of CP, including its functional
semantics~\citep{hirsch-garg-pirouette, cruz-filipe-montesi-core}, location
polymorphism~\citep{alice-or-bob}, and fully out-of-order
execution~\citep{ozone}.

Implementations of CP started with Choral~\citep{choral}, a standalone language
that extends Java with CP features. More recently, HasChor~\citep{haschor}
popularized a library-level approach, providing an EDSL for writing
choreographies, with endpoint projection realized as runtime dispatch. Since
then, library-level CP have been applied to more languages, including
Rust~\citep{conclave}, Clojure~\citep{klor-github},
Elixir~\citep{chorex-github}, and others.

Conditional execution is essential yet challenging in CP, as it must ensure
knowledge of choice~(KoC)~\citep{castagna-knowledge-of-choice}---locations whose
behavior varies across branches are properly informed of the selected
branch/choice. Standalone CP languages typically ensures KoC through
user-provided \emph{select} directives to indicate propagation of choice and a
\emph{merging} phase in endpoint projection to verify their consistency. This
approach offers flexibility for expressing sophisticated choice propagation
patterns, but at the cost of rendering a choreography \emph{unprojectable} if
\emph{select} directives are used inconsistently. Library-level CP takes a
different approach, where KoC is handled by the library and all choreographies
are guaranteed to be projectable. HasChor achieves this by broadcasting the
selected choice to all locations, regardless of their participation in the
branches. To address the inefficiency of HasChor's approach,
\emph{conclaves}~\citep{conclave} have been proposed as a way to constrain the
scope of conditionals, limiting broadcasts to only the relevant locations.

%% file: conclusion.tex
\section{Conclusion}\label{sec:conclusion}

In this paper, we define and study the three variants of freer arrows: freer
pre-arrows, freer arrows, and freer choice arrows. These freer arrows are
amenable to static analysis. To show why static analysis is useful in Haskell,
we conducted a case study on a CP library HasChor. We show that freer arrows
help implement static endpoint projection, endpoint static analysis, and
selective broadcasting, features that were not possible in the original HasChor.

%% file: appendix.tex
\appendix

\section{Obtaining Freer Arrows}\label[appendix]{apx:free-to-freer}

In their paper, \citet{computation-monoids} define the following version of free
arrows:
\begin{lstlisting}[language=Haskell]
data Free a x y where
  Hom :: (x -> y) -> Free a x y
  Comp :: a x z -> Free a z y -> Free a x y
\end{lstlisting}
They further show that \cdh{Free} is a profunctor and a pre-arrow if \cdh{a} is
a profunctor, and \cdh{Free} is a strong profunctor and an arrow if \cdh{a} is a
strong profunctor, given by the following definitions:

\begin{lstlisting}[language=Haskell]
instance Profunctor a => Profunctor (Free a) where
  dimap f g (Hom h)    = Hom (g . h . f)
  dimap f g (Comp x y) = Comp (lmap f x) (rmap g y)

instance Profunctor a => PreArrow (Free a) where
  arr f = Hom f
  c . (Hom f)    = lmap f c
  c . (Comp x y) = Comp x (c . y) 

instance StrongProfunctor a => Arrow (Free a) where
  first (Hom f)    = Hom (first f)
  -- [first'] is a method of strong profunctors.
  first (Comp x y) = Comp (first' x) (first' y)
\end{lstlisting}

However, free arrows are somewhat unsatisfying as they require their first
parameter to be a strong profunctor. We want to make free arrows \emph{freer} by
lifting this restriction. This is important because we want to use freer arrows
with generic algebraic datatypes~(GADTs) like the following:
\begin{lstlisting}[language=Haskell]
-- |- A GADT for stateful effect.
data StateEff :: Type -> Type -> Type -> Type where
  Get :: StateEff s a s
  Put :: StateEff s s s
\end{lstlisting}
Such a GADT is not even a profunctor because one cannot define the \cdh{dimap}
method on it.

This leads to our definition of \emph{freer pre-arrows}. The key idea is
inlining free profunctors in free arrows of \citeauthor{computation-monoids},
similar to how \citet{freer} derive freer monads. This gives you:
\begin{lstlisting}[language=Haskell]
data FreerArrowB e x y where
  Hom :: (x -> y) -> FreerArrowB e x y
  Comp :: (x -> a) ->  (b -> z) -> e a b ->
          FreerArrowB e z y -> FreerArrowB e x y
\end{lstlisting}
From this definition, we take one additional step by fusing the second function
argument \cdh{b -> z} to the function argument of the ``inner''
\cdh{FreerArrow}, which would have type \cdh{z -> c}, where \cdh{c} is a new
existential type.

The definition of \cdh{FreerArrow} can be obtained by inlining free
\emph{strong} profunctors in free arrows and fusing the ``covariance function''
of a \cdh{Comp} constructor with the ``contravariance function'' of the inner
freer arrow, similar to what we did with \cdh{FreerPreArrow}. The definition
of \cdh{FreerChoiceArrow} can be obtained by inlining both free strong
profunctors and free \emph{choice} profunctors.

\section{A Key-Value Server Choreography Using Freer Arrows}\label[appendix]{apx:kvs-arrows}

The following is the arrow version of the HasChor key-value store example with a
client, primary server, and backup server:

\begin{lstlisting}[language=haskell]
discard :: Arrow ar => ar b ()
discard = arr (const ())

asPut :: Request -> Either Request ()
asPut (Put k v) = Left $ Put k v
asPut _ = Right ()

kvs :: Choreo (Kleisli IO) () (Response @ Client)
kvs =
  -- client prepare and send the request
  client `locally0` getRequest >>>
  client ~> primary >>>
  (&&&)
    -- primary handle the request
    (primary `locally` handleRequest)
    -- disparate behavior based on the request
    (cond' primary (arr asPut) $
      (|||)
        -- propogate Put k v
        (arr wrap >>>
        primary ~> backup >>>
        backup `locally` handleRequest >>>
        discard)
        -- do nothing with Get k
        discard
    ) >>>
  arr fst >>>
  primary ~> client
\end{lstlisting}

The operator \cdh{(&&&)} is used to establish that two things happen in
parallel: the primary server handles the request, and the primary server
forwards the request to the backup server to handle if the request is some
\cdh{Put k v}. The combinator \cdh{cond'} branches on the result of a local
computation:

\begin{lstlisting}[language=Haskell]
cond' :: (Show x, Read x, KnownSymbol l)
      => Proxy l
      -> ar b x -> Choreo ar x a
      -> Choreo ar (b @ l) a
\end{lstlisting}

Here, the result of the local computation \cdh{arr asPut} is \cdh{Left} of the
original request if it's a \cdh{Put} request and \cdh{Right} otherwise. This
choice is broadcast to the participants of the continuation choreography so that
we can branch using \cdh{(|||)}. Each branch of \cdh{(|||)} receives its data
unwrapped, \ie~it has no annotation \cdh{@ l}, since it was already broadcast to
each endpoint. This means that we have to \cdh{wrap} the data to use it to be
able to run a local computation on it or send it.

The result of \cdh{cond'} here is \cdh{()} which we can discard by taking only
the result of the left branch of \cdh{(&&&)} using \cdh{arr fst} before sending
the result back to the client.

\section{Extensible Effects With Freer Arrows}\label[appendix]{apx:composable-eff}

The achieve extensible effects, it's useful to be able to construct compound
effects which compose multiple effects together. This approach is used for
extensible effects with freer monads~\citet{freer}. The \cdh{Sum2} datatype
gives a simple way to do just that; it lets use compose two types of effects as
a sum with constructors \cdh{Inl2} and \cdh{Inr2} resembling the \cdh{Left} and
\cdh{Right} constructors of \cdh{Either}. The second class \cdh{Inj2} gives us a
more flexible way of representing the inclusion of one effect in another by
providing a function for injecting an event of one type into the other. With
\cdh{Inj2}, we can write code that is polymorphic over any effect signature that
contains the desired effect.

\begin{lstlisting}[language=Haskell]
-- |- The [Sum2] datatype
data Sum2 (l :: Type -> Type -> Type)
          (r :: Type -> Type -> Type) a b where
  Inl2 :: l a b -> Sum2 l r a b
  Inr2 :: r a b -> Sum2 l r a b

-- |- An injection relation between effects.
class Inj2 (a :: Type -> Type -> Type)
           (b :: Type -> Type -> Type) where
  inj2 :: a x y -> b x y

-- |- Automatic inference using typeclass resolution.  
instance Inj2 l l where
  inj2 = id

instance Inj2 l (Sum2 l r) where 
  inj2 = Inl2

instance Inj2 r r' => Inj2 r (Sum2 l r') where
  inj2 = Inr2 . inj2

-- | A more generalized instance showing that freer
-- arrows are an ArrowState.
instance Inj2 (StateEff s) e =>
         ArrowState s (FreerArrow e) where
  get = embed $ inj2 Get
  put = embed $ inj2 Put
\end{lstlisting}

\section{Datatype for Elgot Iteration}\label[appendix]{apx:elgot}

We can reify the Elgot iteration~\citep{elgot, elgot-theories} using the
following datatype:
\begin{lstlisting}[language=haskell]
data Elgot f (e :: Type -> Type -> Type) x y where
  Elgot :: f e x (Either z x) ->
           f e z y -> Elgot f e x y
\end{lstlisting}
The \cdh{Elgot} datatype is parameterized by four parameters, an \cdh{f},
an \cdh{e} of kind {Type -> Type -> Type}, an \cdh{x}, and a \cdh{y}. It contains only
one constructor, also named \cdh{Elgot}. The constructor takes two arguments: a
``loop body'' of type \cdh{f e x (Either z x)} and a continuation of type \cdh{f
e z y}. The loop body takes an input type \cdh{x} and returns a sum
type \cdh{Either z x}, with a left value of type \cdh{z} indicating that the
loop has ended, and a right value of type \cdh{x}, which is the same type as the
input type \cdh{x}, indicating that the loop should continue.

We can interpret \cdh{Elgot} by combining it with a freer choice arrow:
\begin{lstlisting}[language=haskell]
interp :: ArrowChoice arr =>
  (f e :-> arr) -> Elgot f e x y -> arr x y
interp h (Elgot l k) =
  let l' = h l in
  let k' = h k in
  let go = l' >>> k' ||| go in go
\end{lstlisting}

The following function demonstrates that we can define a countdown function
using \cdh{Elgot} and \cdh{FreerChoiceArrow}:
\begin{lstlisting}[language=haskell]
countA :: Elgot FreerChoiceArrow
                (StateEff Int) Int Int
countA =
  let go :: FreerChoiceArrow (StateEff Int)
                             Int
                             (Either Int Int)
      go = get >>>
           arr (\n -> if n == 0 then Left n
                                else Right n) >>>
           right (lmap (\x -> x - 1) put) in
    Elgot go id
\end{lstlisting}